\documentclass{camera}
\usepackage{graphicx}  

\begin{document}

%
\title{Ultra High Energy Cosmic Rays}

%
\author{Pasquale Blasi}

%
\organization{INAF/Osservatorio Astrofisico di Arcetri}

\maketitle

\begin{abstract}
The origin of the particles in the highest energy end of the cosmic
ray spectrum is discussed in the context of the wider problem of the
origin of the whole cosmic radiation as observed at
the Earth. In particular we focus our attention on the
acceleration problem and on the transition from galactic to
extragalactic cosmic rays.
\end{abstract}

\section{Introduction}

There are mainly two reasons for the attention shown by the
community for the origin of the highest energy cosmic
ray spectrum, say above $10^{19}$ eV: on the one hand most
acceleration processes need to be pushed to their extreme (or
beyond) in the attempt to reach such energies. On the other hand,
even in the presence of numerous nearby sources, the spectrum is
expected to show a suppression, known as the GZK feature (Greisen
1966; Zatsepin \& Kuzmin 1966) above the threshold for photopion
production on the photons of the cosmic microwave background (CMB).
The suppression should become evident around $\sim 10^{20}$ eV. Here
we review some recent ideas put forward to address both points
mentioned above.

The acceleration of cosmic rays to the observed energies is a
problem even in supernova remnants, as possible sources of galactic
cosmic rays, therefore it is natural to try and learn from that
instance. As for the presence of the GZK feature we summarize
the results of current experiments and try to address the issue of
what future experiments could be expected to observe. However, in
order to understand the origin of extragalactic cosmic rays and more
specifically of UHECRs, it is crucial to understand which cosmic
rays are in fact extragalactic. The two main lines of thought in
this respect will be summarized and discussed: in the so-called {\it
ankle scenario} (Hillas 2005; Wibig \& wolfendale 2005) the
transition takes place around $10^{19}$ eV where a steep galactic
spectrum encounters the flat spectrum of extragalactic cosmic rays.
In the {\it dip scenario} (Berezinsky et al. 2002, 2005) the
transition takes place at energies roughly one order of magnitude
lower. The case of a mixed chemical composition of extragalactic
cosmic rays (Allard et al. 2005a; Allard et al. 2005b) will also be
discussed.

The paper is structured as follows: in \S \ref{sec:acceleration} we
discuss the problem of acceleration and some recent findings on
magnetic field amplification in the acceleration region. In \S
\ref{sec:trans} we discuss the current ideas on the transition
between galactic and extragalactic cosmic rays. In \S \ref{sec:GZK}
we describe the current status of the observations of the end
of the cosmic ray spectrum and of the anisotropies in such energy
range. We conclude in \S \ref{sec:concl}.

\section{Recent ideas on acceleration of Cosmic Rays}
\label{sec:acceleration}

While it is often argued that the known acceleration mechanisms have
serious problems in achieving energies in excess of $10^{20}$ eV, it
is sometimes not recognized that supernova remnants (SNRs), the most
serious candidates as accelerators of galactic CRs, have the same
problems in accelerating particles above the knee ($\sim 10^{15}$
eV). Here we start from this analogy to gather information on
the source of the problem and try to assess its possible solutions.

In the case of SNRs, acceleration in assumed to take place at the
shock front associated with the supersonic motion of the expanding
shell. Particles are energized through diffusive acceleration {\it a
la Fermi}. The acceleration time is given by the well known
expression $\tau_{acc}=\frac{3}{u_1-u_2}\left[ \frac{D_1}{u_1} +
\frac{D_2}{u_2} \right]$, where the subscripts $1$ and $2$ refer to
the upstream and downstream region respectively. For a strong shock
$u_1/u_2\approx 4$ and if we assume that the diffusion coefficient
is not changed dramatically at the shock ($D_1\approx D_2$), the
acceleration time can be written as:
\begin{equation}
\tau_{acc} (E) = \frac{20}{3} \frac{D_1(E)}{u_1^2}.
\end{equation}
If the supernova shell is expanding in the interstellar medium, it
is reasonable to take as a diffusion coefficient a form that is
often used to describe the propagation of cosmic rays in the Galaxy,
$D(E)=A E^\alpha$ with $A=3\times 10^{27}cm^2 s^{-1}$ ($A=3\times
10^{29}cm^2 s^{-1}$) for $\alpha=0.6$ ($\alpha=0.3$) and $E$ is the
particle energy in units of GeV. The maximum energy of the
accelerated particles can be estimated by comparing the acceleration
time with the age of the remnant. For both choices of the diffusion
coefficient given above, the maximum energy turns out to be of
in the GeV range for a remnant of $\sim 1000$ years: the magnetic
scattering provided by the ISM is insufficient to warrant the
acceleration of cosmic rays to the observed energies. This
conclusion is not appreciably changed by taking into account the
mild compression of the perpendicular components of the magnetic
field at the shock front. A crucial consequence of this finding is
that the mechanism of diffusive particle acceleration at supernova
shocks is efficient only if additional scattering exists close to
the shock, either because the circumstellar material provides it, or
because the accelerated particles generate a larger magnetic field
$\delta B$ from the background field $B$, through streaming
instability. The possibility of magnetic field amplification was
already discussed in (Bell 1978; Lagage \& Cesarsky 1983a,b), where
its consequences on the maximum achievable energy were also
evaluated. The conclusion of Lagage \& Cesarsky (1983a,b) was that
shocks in SNRs could accelerate cosmic rays up to $\sim 10^4$ GeV,
below the knee, if the amplification results in $\delta B/B\sim 1$
(on all spatial scales) and the diffusion coefficient has the form
of the Bohm diffusion, $D(E)=10^{23} E(GeV) \delta B_\mu^{-1} cm^2
s^{-1}$, where $\delta B_\mu$ is the amplified magnetic field in
$\mu G$. Using this diffusion coefficient, the maximum energy
evaluated as above turns out to be $E_{max}\sim (1-4)\times 10^4$
GeV for $u_1=5000-10000 \rm km s^{-1}$, $\delta B_\mu=1$ and a
supernova age of 1000 years. Slightly larger (or smaller) values of
$E_{max}$ can be obtained for slightly different values of the
parameters.

There are two main issues of physical relevance here, whose validity
goes well beyond the specific case of SNRs: 1) the nature of the
instabilities that lead to the magnetic field amplification and 2)
the saturation of the amplified magnetic field. Both affect the
maximum energy of the particles with respect to the values quoted
above.

The streaming of the accelerated particles upstream of the shock at
super-Alfvenic speed leads to streaming instability, as already
discussed by Bell (1978) (see also references therein). The maximum
values of the amplified magnetic field $\delta B$ is however not
limited by the value of the background field, but by the value
\begin{equation}
\delta B = B \left[ 2 M_A \frac{P_{CR}}{\rho u^2} \right]^{1/2},
\label{eq:deltaB}
\end{equation}
where $\rho u^2$ is the ram pressure of the inflowing fluid in the
upstream region, $P_{CR}$ is the pressure in the form of accelerated
particles, and $M_A = u\sqrt{4\pi \rho}/B$ is the Alfvenic Mach
number of the upstream fluid. All these results are obtained in the context
of the quasi-linear theory and should in principle been used only
for $\delta B/B\ll 1$, while they predict $\delta B/B\gg 1$,
therefore these conclusions should be used with much care. The waves
that turn nonlinear within this approach are Alfven waves.

Recently Bell and Lucek (2000) have presented evidence for magnetic
fields much larger than those given by Eq. \ref{eq:deltaB}. The
interpretation of this result appears in (Bell 2004) where a new
non-alfvenic purely growing mode is identified. Quasi-linear theory
suggests that the saturation of the growth of these waves should
take place at
\begin{equation}
\delta B = B M_A \sqrt{\frac{u}{c}\frac{P_{CR}}{\rho u^2}},
\end{equation}
where the symbols have the same meaning as above. This saturation
level seems to be confirmed by hybrid simulations presented in (Bell
2004). For typical parameters of a SNR and assuming that an
appreciable fraction of the kinetic pressure is transformed into
cosmic rays ($\rho u^2 \approx P_{CR}$), one can predict $\delta
B/B\sim 500$ (versus $\delta B/B\sim 20$ in the previous case). If
one assumes Bohm diffusion, this translates to correspondingly
higher values of the maximum energy of accelerated particles: for
$\delta B/B\sim 500$ one has $E_{max}\sim (0.5-2)\times 10^7$ GeV
(assuming Bohm diffusion).

The condition $\rho u^2 \approx P_{CR}$ is typically found to be a
consequence of the nonlinear effects in particle acceleration,
namely the effects due to the dynamical reaction of the accelerated
particles (see Malkov \& Drury (2001) and Blasi (these proceedings)
for a review). This reaction, which leads to the so-called {\it
cosmic ray modification} of shocks, has three important
phenomenological consequences: 1) creation of a cosmic ray precursor
which is also responsible for a concave energy spectrum (non power
law); 2) large efficiency for particle acceleration; 3) suppression
of the plasma heating in cosmic ray modified shocks.

A unified picture of nonlinear particle acceleration at shocks with
self-generation of scattering has recently been presented by Amato
\& Blasi (2005, 2006). An important effect of the shock modification
is that while the amplification of the magnetic field leads to an
increase of the maximum achievable momentum, the precursor (slowing
down of the upstream fluid and spatial variation of the magnetic
field) lead to a somewhat lower value of the maximum energy (Blasi,
Amato \& Caprioli 2006).

\begin{figure}
\resizebox{\hsize}{!}{
\includegraphics{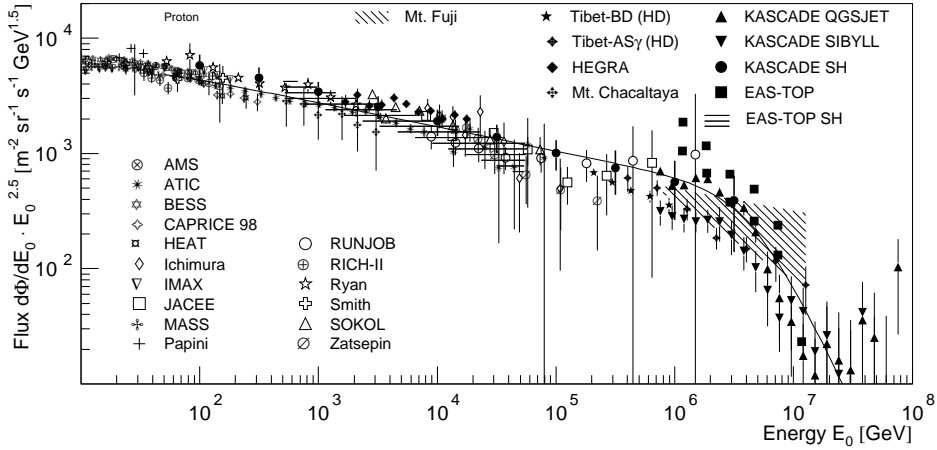}
} \caption{Spectrum of protons} 
\label{fig:KASCADE} 
\end{figure}

The observational picture provided by the KASCADE data is extremely
important to identify the sources of cosmic rays in the Galaxy. In
Fig. \ref{fig:KASCADE} we plot the spectrum of protons as measured
by KASCADE and other experiments (Hoerandel 2005). The proton knee
is clearly visible, while the spectrum extends up to at least $10^7$
GeV, although it is not clear whether the steepening above the knee
is a cutoff due to an approaching $E_{max}$ or rather a slope change
either due to acceleration or to propagation in the Galaxy. The
spectrum of helium nuclei appears to extend to slightly higher
energies, as it could be expected in a rigidity dependent model of
acceleration. In this picture the knee in the iron component would
be expected to be at energy $E_k^{Fe} = Z E_k^p \approx 8\times
10^{16}$ eV, while the spectrum would probably extend up to
$E_{max}^{Fe} = Z E_{max}^p \approx 2\times 10^{17}$ eV. The
spectrum of iron nuclei is however not observed in a reliable way at
this time, therefore this should be considered as a phenomenological
conclusion. But a very important one: in fact it is difficult to
imagine more energetic particles above the maximum energy of iron
nuclei, being originated inside the Galaxy. Therefore, current
observations, together with the most recent theoretical
understanding of the acceleration processes at shock fronts hint to
the fact that the galactic component of cosmic rays should end
around $\sim 2\times 10^{17}$ eV.

These observational findings clearly indicate that if SNRs are in
fact the sources of galactic cosmic rays, then substantial magnetic
field amplification should take place at the shock. Independent
evidences for such amplification comes from Chandra X-ray
observations of the X-ray rim of several SNRs, produced as 
synchrotron emission of
relativistic electrons accelerated at the shock front. It has been
pointed out that the spatial extension of these regions is
compatible with magnetic fields of the order of $\sim 100-300\mu G$
and not with the typical fields in the interstellar medium (Warren et
al. 2005).

The example of SNRs illustrated above shows that in order to assess
the ability for a class of sources to accelerate particles to a
given energy it is absolutely crucial to take into account the
nonlinear processes which in fact make the acceleration possible and
efficient. This applies equally well to the putative sources of
UHECRs, such as active galactic nuclei (AGNs) or gamma ray bursts
(GRBs) could be. Unfortunately, in both these cases the acceleration
is likely to involve relativistic plasma motions, for which the
investigation of the shock acceleration process has mostly been
studied in the linear regime. An extension of the nonlinear theory
to these cases is of the highest importance.

The most valuable lesson that can be probably learned from the case of
SNRs is that in estimating the maximum energy achievable in a cosmic
accelerator, the value of the local ambient magnetic field might not
be the physically relevant quantity. The self-generated amplified
magnetic field in the vicinity of the source can in fact wildly exceed
the ambient field, thereby enhancing the maximum energy as inferred
from Hillas-like plots. The topology of the turbulent field is also
extremely important: the same rms field leads to different diffusion
properties of cosmic rays for different topologies. For self-generated
fields, the topology is determined by the particles which are being
accelerated, which makes it extremely difficult to carry out realistic
self-consistent calculations.

\section{The transition from galactic to extragalactic cosmic rays}
\label{sec:trans}

It has long been thought that the ankle, at $\sim (0.5-1)\times 
10^{19}$ eV is a
feature arising from the intersection of a steeply falling galactic
spectrum of cosmic rays with a flatter spectrum of extragalactic
cosmic rays. This view implies that the galactic cosmic ray spectrum
should extend to energies in excess of $\sim 10^{19}$ eV. However
the arguments illustrated in Sec. \ref{sec:acceleration} suggest an
early end of the galactic component, between $10^{17}$ and $10^{18}$
eV, although as we stressed, this inference is not purely based on
observations. Where does the transition actually take place?
\begin{figure}[ht]
\includegraphics[width=5.6cm,clip]{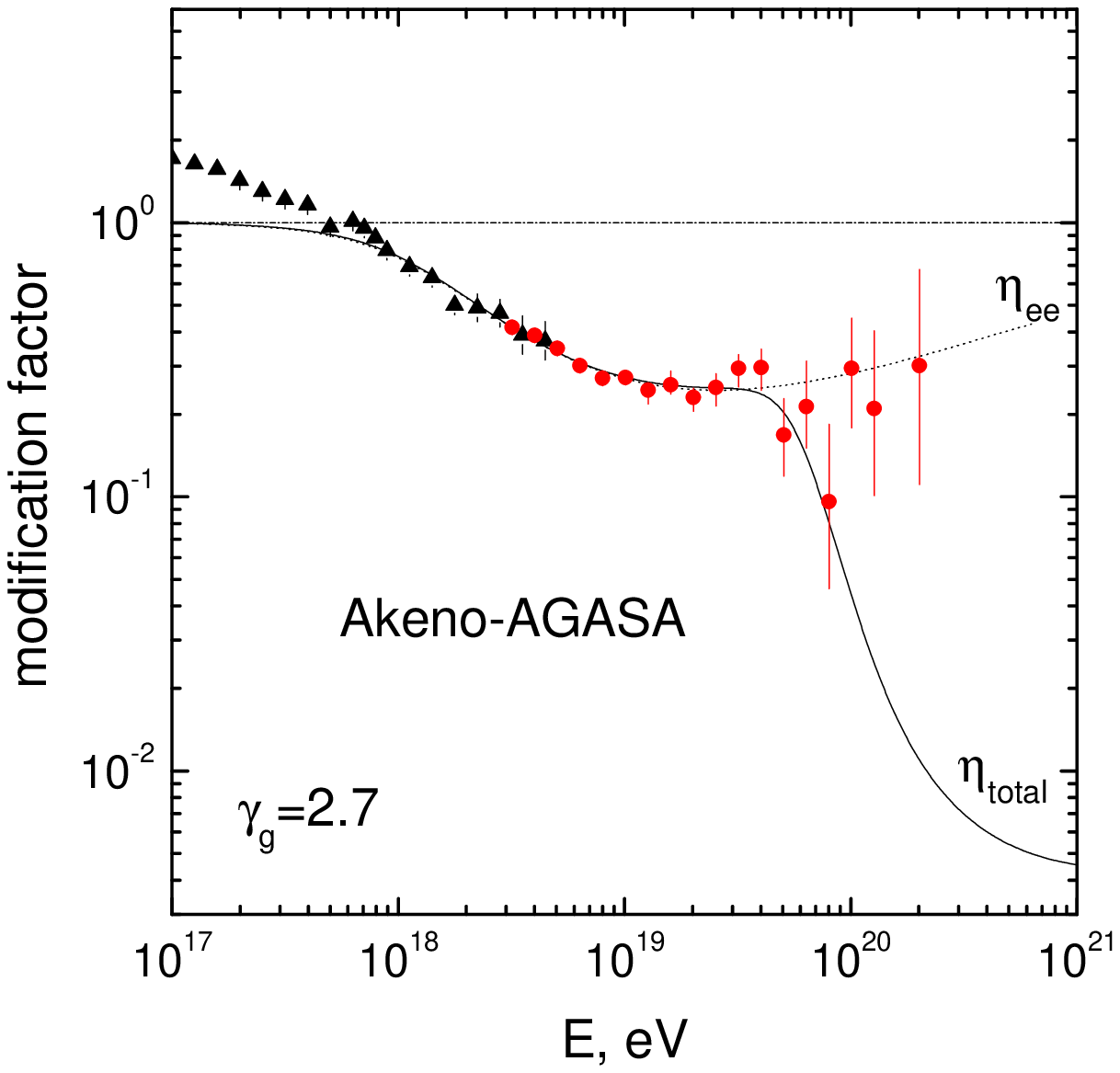}
\includegraphics[width=5.6cm,clip]{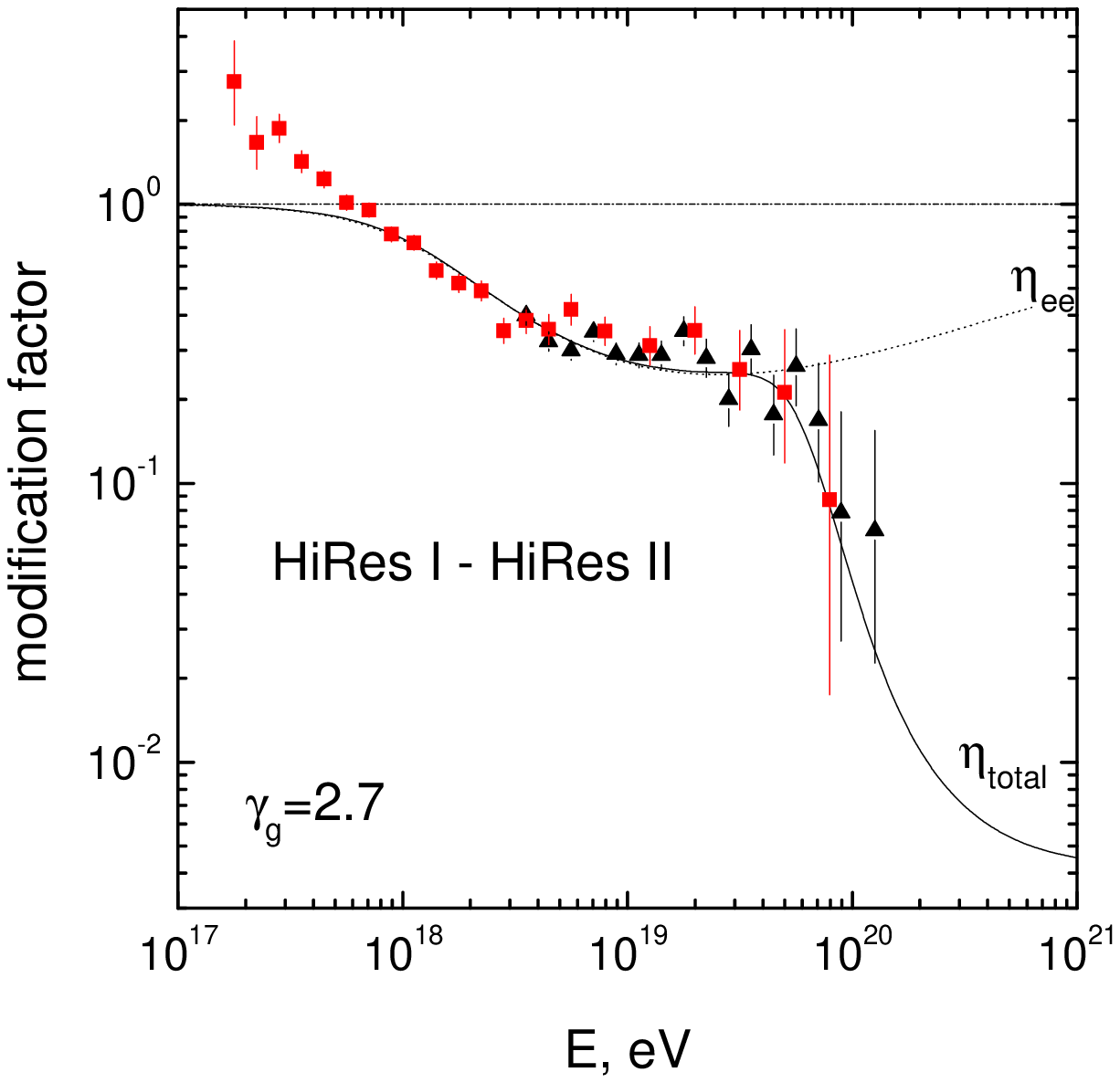}
\includegraphics[width=5.6cm,clip]{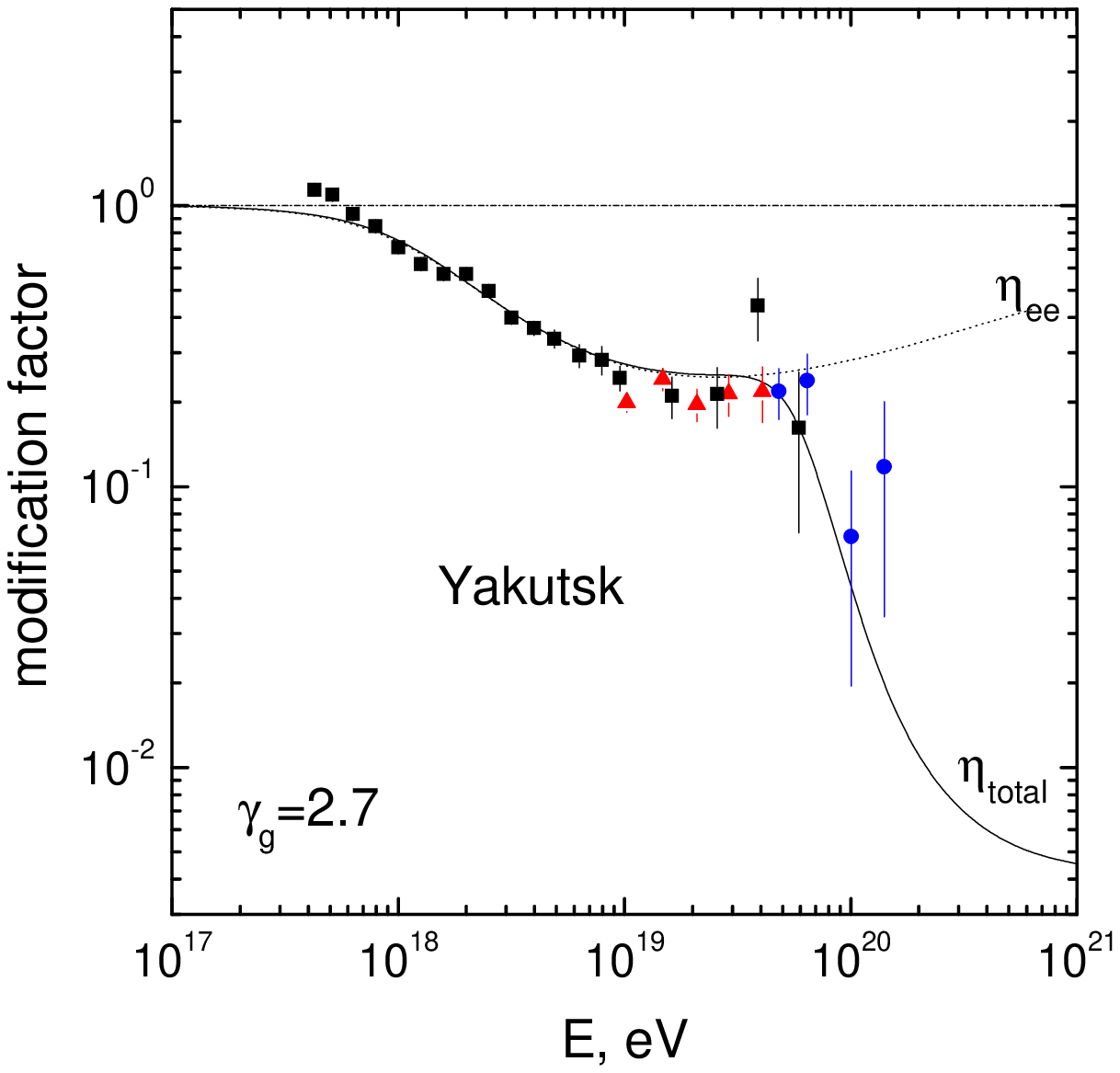}
\hspace{12mm}
\includegraphics[width=5.6cm,clip]{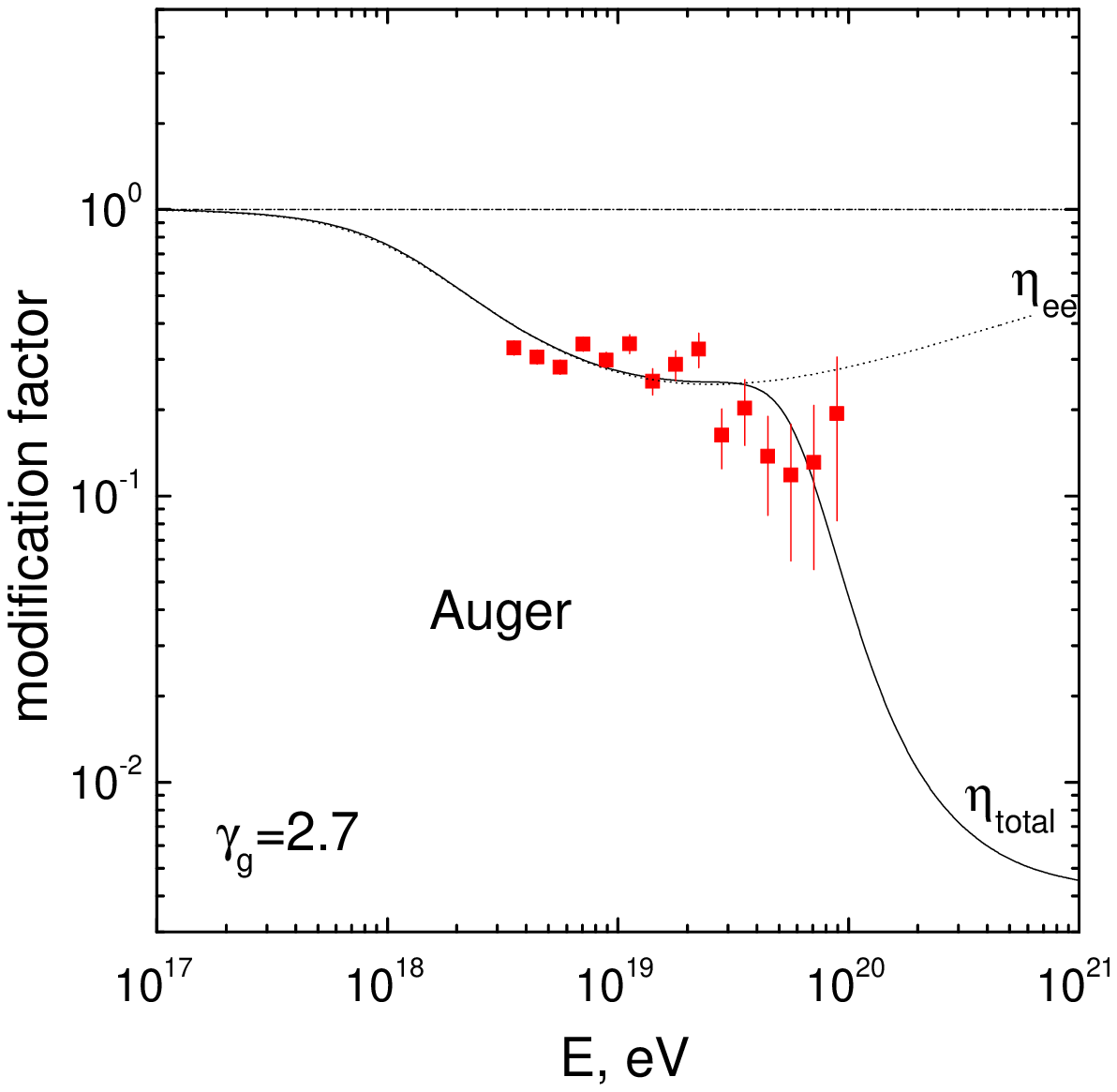}
\caption{\label{fig:dip} Predicted dip in comparison with AGASA,
HiRes, Yakutsk and Auger data.}
\end{figure}
As pointed out by Berezinsky et al. (2002, 2005), a dip appears 
in the spectrum of
extragalactic cosmic rays at energy $\sim 3\times 10^{18}$ eV, as
due to the combination of adiabatic losses (expansion of the
universe) and $e^\pm$ pair production. This dip was shown to fit
very well the observed spectra for all the relevant experiments, as
also shown in Fig. \ref{fig:dip}. A very important point to make is
that the position of the dip is independent of astrophysical
details, and is fixed by the rates of adiabatic losses and pair
production. The low energy part of the dip fits what is currently
named the {\it second knee}. Below the second knee the predicted spectrum
flattens and drops below the flux of galactic cosmic rays. This low
energy part depends to some extent on the mean distance between
sources and on the magnetic field value and topology in the
intergalactic medium (Aloisio \& Berezinsky 2005; Lemoine 2005). In
this scenario the transition between galactic and extragalactic
cosmic rays takes place somewhere between $10^{17}$ eV and $10^{18}$
eV, but it remains true that in the transition region a steep
galactic spectrum encounters a flatter extragalactic spectrum
(Berezinsky et al. 2006), as in the {\it ankle} scenario.

Both possibilities are currently viable and have positive and
negative aspects. The injection spectrum required to fit the data in
the ankle scenario is as flat as $E^{-\alpha}$, with $\alpha\sim
2-2.4$, which is tantalizingly close to the results expected for
shock acceleration, while in the dip scenario the injection spectrum
has $\alpha=2.6-2.7$. However, it was pointed out in (Kachelriess \&
Semikoz 2005) that the superposition of flat spectra with different
maximum energies naturally provides a good fit to the data without
requiring a steep injection spectrum. In any case one should always
keep in mind that $\alpha\sim 2$ is more likely a number we got used
to rather than a value than Nature prefers, as discussed by
Berezinsky et al. (2006): many physical situations contribute to
steepen the spectrum with respect to the canonical case $\alpha\sim
2$.

From the point of view of the chemical composition the two models
differ the most: in the ankle scenario (Wibig \& Wolfendale 2005), 
the galactic cosmic rays extend to $>10^{19}$ eV and are mainly iron
nuclei, while the dip
scenario requires that CRs with energy above $\sim 10^{19}$ eV are
mostly protons (with no more than $\sim 15\%$ contamination of
helium) and that the proton dominated extragalactic component is
important down to energies around $\sim 10^{18}$ eV. 
The dip scenario would naturally account for the $\sim
10\%$ proton abundance observed by Akeno at $10^{17}$ eV. The
difference in the prediction of the chemical composition of CRs also
represents the most striking prediction of the models and the tool
to possibly discriminate between them.

An important aspect of the dip scenario is that it provides a
description of the transition from galactic to extragalactic cosmic
rays which is consistent with the KASCADE observations. On the other
hand, the ankle scenario requires that galactic sources should be
able to accelerate cosmic rays up to $\sim 10^{19}$ eV, which
appears rather challenging on the basis of current knowledge of
acceleration processes in galactic sources.

In addition to the two scenarios discussed above, there is a third
one, based on the possibility that the chemical composition
at the source is contaminated by nuclei heavier than hydrogen
(Allard et al. 2005a; Allard et al. 2005b). The propagation of these
elements and their fragmentation in the cosmic photon background
determine a rather complex energy dependent chemical composition at
the Earth, which depends somewhat on the assumptions on the injection
spectra and relative abundancies in the sources. In this model the 
transition between galactic and
extragalactic cosmic rays takes place at $\sim 2\times 10^{18}$ eV,
not very different from but slightly higher than the prediction of 
the dip scenario. As a
consequence, this model also appears to agree with the fact that the
galactic spectrum should end at energies $10^{17}-10^{18}$ eV. 
However, the mixed composition model requires a rather
flat injection spectrum and predicts that the chemical composition
in the transition region has a strong iron and helium contamination.
In (Allard et al. 2005b) the authors argue that the model is in good
agreement with the elongation rate observed by different
experiments. This conclusion is reached by adopting a specific
recipe for the chemical abundance in the accelerator and then
studying the dependence of the results on the initial assumptions.
Unfortunately the data on the chemical composition as derived from
the elongation rate (or any other method so far) are rather heavily
affected by uncertainties in the interaction models and a final
confirmation of one or the other of these models will require
dedicated measurements of the chemical composition in the transition
region and possibly a better understanding of the physics entering
the description of the interactions in the atmosphere in this energy
region.

\section{Spectrum and anisotropies of UHECRs}
\label{sec:GZK}

The discussion on the spectrum of UHECRs is often concentrated on 
the presence or absence of the GZK feature, a flux suppression at
energies in excess of $10^{20}$ eV resulting from photopion
production interactions of protons with the cosmic microwave
background. The theoretical prediction of this part of the spectrum
is extremely uncertain, being dependent on the injection spectrum,
the distribution and spatial density of the sources and the strength
and topology of the intergalactic magnetic field. The search for
this feature has given inconclusive results so far, mainly due to
the very low statistics of detected events. From the statistical
point of view, the most significant data are those collected by
AGASA, HiRes and the Pierre Auger Observatory. AGASA and HiRes, with
comparable exposures, have results which are discrepant in the
highest energy part: the spectrum of AGASA does not show the GZK
suppression, while HiRes spectrum has a pronounced GZK feature. The
numerical simulations of De Marco, Blasi \& Olinto (2003) showed
however that, given the small number of collected events, the
discrepancy is in fact at $\sim 2-3\sigma$ level, being further reduced if
the offset in the overall normalization of the spectra is attributed
to a systematic error in the energy determination. A systematic
error of $\sim 30\%$ would in fact make the experiments to
reasonably agree with each other. It is important to stress that the
simulation of the propagation of UHECRs was carried out by De Marco,
Blasi \& Olinto (2003) in the case of a truly continuous
distribution of the sources (no point sources with finite density).

The most recent measurement of the spectrum by the Pierre Auger
Observatory (Sommers et al., 2005) is in closer agreement with
the HiRes results, although again no conclusive evidence for the
absence of the GZK feature can be claimed so far.

An important piece of information on the sources of UHECRs and
indirectly even on their injection spectrum, can be inferred from
small scale anisotropies (SSA), that signal for the presence of
point sources. The SSA can be measured through the effect they have
on the two point correlation function of the arrival directions of
the detected events, and clearly they provide information on point
sources only if the magnetic field in the propagation volume is not
significant (Blasi \& De Marco, 2003). The two point correlation
function of the AGASA data, when compared with the simulated events
provides an estimate of the source number density of $\sim
10^{-5}\rm Mpc^{-3}$ (Blasi \& De Marco, 2003), with a very large
uncertainty (about one order of magnitude on both sides) that
results from the limited statistics of events above $4\times
10^{19}$ eV, where the analysis should be carried out in order to
avoid (or limit) the effects of the galactic magnetic field.

If we use the best fit for the source density $n_s \sim 10^{-5}\rm
Mpc^{-3}$ (Blasi \& De Marco, 2003) and determine the simulated
spectrum of UHECRs at the Earth, we can see that AGASA small scale
anisotropies appear to be inconsistent with the spectrum measured by
the same experiment at the level of $\sim 5\sigma$ (De Marco, Blasi
\& Olinto 2006a). Given the large error in the determination of $n_s$
this result cannot be taken too seriously, but it certainly hints to
the possibility that the SSA observed by AGASA may be a statistical
fluctuation. Its statistical significance has in fact been shown to
be rather weak (Finley \& Westerhoff 2004) and dependent upon the
choice of the binning angle for the arrival directions (Finley \&
Westerhoff 2004; De Marco, Blasi \& Olinto 2006a). Moreover, the
technique of a combined study of the SSA and spectrum of UHECRs
appears to be promising in the perspective of the wealth of data
that the Pierre Auger Observatory will provide us with.

The investigation of the SSA through the two point correlation
function is however likely to be difficult
even with the results that will come out of the Auger South
observatory. In (De Marco, Blasi \& Olinto 2006b) the simulations of
propagation from point sources were repeated for the expected Auger
statistics of events: the results suggest that although it will be
easy to distinguish the case of point sources from the case of a
purely homogeneous distribution, it will not be easy to achieve a
good resolving power between different values of the source density
$n_s$. Accounting for the galactic magnetic field and luminosity
function of the sources contributes to emphasize this difficulty.
On the other hand, the SSA should also result in the appearance of
{\it hot spots} in the UHECR sky, so that the search for the sources
can proceed though alternative routes, such as the identification of
counterparts or the cross-correlation of arrival directions with the
positions of sources in given catalogs. The power of these analyses is
expected to be much larger when both Auger South and Auger North will
be available. 

\section{Conclusions}
\label{sec:concl}

We discussed three of the crucial issues in the investigation of the
origin of UHECRs: 1) Particle acceleration in astrophysical sources;
2) the transition from galactic to extragalactic cosmic rays; 3) the
spectrum and anisotropies of UHECRs.

The theoretical investigation of particle acceleration at shock
fronts is recently experiencing a boost, due to some interesting
calculations of the nonlinear effects of particle acceleration (e.g.
dynamical reaction of the accelerated particles and self-generation
of scattering) and some observational results on the spectra of
single chemical components (e.g. measurements from KASCADE). These 
lines of investigation are leading to some
consensus on the fact that protons may be accelerated by sources in
the Galaxy up to energies $\sim 10^{16}$ eV. A rigidity
dependent argument would imply that the spectrum of iron extends to
energy $\sim 3\times 10^{17}$ eV. The flux of galactic cosmic rays
should rapidly decay above this energy, so that a transition to
extragalactic cosmic rays is expected to take place at energy $\sim
10^{18}$ eV. This conclusion is at odds with the traditional
interpretation of the ankle (Wibig and Wolfendale 2005) as the 
transition feature, while it
appears to agree with the prediction of the dip scenario (Berezinsky
et al. 2006). A similar energy for the transition energy is however 
inferred in the case of a
mixed composition of extragalactic cosmic rays, as discussed in
(Allard et al. 2005a; Allard et al. 2005b). Whether the dip or the
mixed composition scenario best describes the observations will most
likely be decided when an accurate measurement of the chemical
composition between $10^{17}$ and $10^{19}$ eV will become
available.

At larger energies, a crucial issue is represented by the
measurement of the spectrum at energies around $10^{20}$ eV, where
the GZK suppression has long been searched and missed. AGASA and HiRes spectra
are in a mild contradiction which may well be due to statistical
fluctuations and a systematic error in the energy determination with
the two different techniques used. The Pierre Auger Observatory
should soon settle this issue.

At sufficiently high energies the magnetic field of the Galaxy is
not expected to bend the trajectories of high energy particles by
more than the angular resolution of the operating cosmic ray
experiments. This may lead to the identification of small scale
anisotropies, flagging the presence of discrete sources of UHECRs. A
signal of this type was found by AGASA, though its statistical
significance was later questioned (Finley \& Westerhoff 2004; De
Marco, Blasi \& Olinto 2006b). The level of SSA detected by AGASA
would correspond to a source density $\sim 10^{-5}\rm Mpc^{-3}$
(Blasi and De Marco 2004). This density would roughly correspond to the density
of active galactic nuclei. This has inspired numerous searches for
correlations between the directions of arrival of UHECRs and the
locations of AGNs. Despite several claims of positive correlations,
a definite answer is still missing.

\section*{Acknowledgments}

The author is very grateful to R. Aloisio, E. Amato, V. Berezinsky,
and D. De Marco for ongoing collaboration on topics of relevance for
this review.

\end{document}